\documentclass[aps,pra,reprint,nofootinbib,showpacs,floatfix]{revtex4-2}
\usepackage{amsfonts}
\usepackage{amsmath,amssymb}
\usepackage{physics}
\usepackage{tikz}
\usepackage{qcircuit}
\usepackage{graphicx}
\usepackage{blkarray}
\usepackage{float}
\usepackage{subfigure}
\usepackage{bm}
\usepackage{xcolor}
\usepackage{hyperref}
\usepackage{tikz}
\usetikzlibrary{arrows, positioning, shapes, decorations.pathreplacing}
\hypersetup{
     colorlinks   = true,
     citecolor    = blue
}
\usepackage{multibib}

\def\be{\begin{equation}}
\def\ee{\end{equation}}
\def\bea{\begin{eqnarray}}
\def\eea{\end{eqnarray}}
\begin{document}
	\title{Practical Quantum Clock Synchronization Using Weak Coherent Pulses}
	\author{Noah Crum}
	\email{ncrum@vols.utk.edu}
	\affiliation{Department of Physics and Astronomy, The University of Tennessee, Knoxville, TN 37996-1200, USA}
	\author{Md Mehdi Hassan}
	\email{mhassa11@vols.utk.edu}
	\affiliation{Department of Physics and Astronomy, The University of Tennessee, Knoxville, TN 37996-1200, USA}
	
	\author{George Siopsis}
	\email{siopsis@tennessee.edu}
	\affiliation{Department of Physics and Astronomy, The University of Tennessee, Knoxville, TN 37996-1200, USA}
	\date{\today}

\begin{abstract}
Establishing and maintaining a common time reference across spatially separated devices is a prerequisite for networked quantum experiments and secure communications.
Classical two–way timing protocols such as Network Time Protocol (NTP) or Precision Time Protocol (PTP) are vulnerable to asymmetric channel delays and cannot provide the picosecond‑level precision demanded by quantum repeater networks.
We propose and numerically evaluate a quantum‑enhanced clock synchronization protocol based on attenuated weak coherent pulses (WCPs) and bidirectional Hong–Ou–Mandel (HOM) interferometry.
Our simulations assume telecom‑band photons (1550~nm) with a temporal width of $10.0$~ns, a repetition rate of $f=10$~MHz, effective mean photon number $\mu=1.0$, detector efficiency $\eta=85\%$, detector timing jitter of $150$~ps and channel loss $0.2$~dB/km.
We simulate that sub‑nanosecond clock‑offset accuracy and precision can be achieved under these operating conditions.
This work demonstrates that high‑repetition‑rate WCPs combined with HOM interference can provide flexible and secure quantum clock synchronization at sub‑nanosecond precision.
\end{abstract}
	\maketitle

\section{Introduction}

Clock synchronization is the process of establishing a consistent time scale across multiple devices or systems. It is a fundamental requirement for many technological applications, ranging from telecommunications and navigation systems to scientific experiments and financial transactions. Accurate clock synchronization is particularly important in modern communication networks, where the precise timing of signals is essential for maintaining network efficiency and reliability and is imperative for entanglement swapping procedures \cite{Burenkov_23, Alshowkan_22}.

Current popular classical approaches to the problem of clock synchronization utilize the timing of signal propagation to interrogate clock offsets. Techniques such as Network Time Protocol (NTP) \cite{NTP} and Precision Time Protocol (PTP) \cite{PTP} are susceptible to asymmetric delays of the synchronization signal \cite{Ullmann_delays, naval_spoofing, Quan_2024}. These asymmetric delays can be the byproduct of an ill-intentioned adversary hoping to introduce inaccuracies in timing to a network construction. Recently, White Rabbit synchronization is being leveraged as a classical method for frequency and time transfer \cite{gilligan_2020}. We propose a protocol relying on attenuated weak coherent pulses (WCPs) and the Hong-Ou-Mandel (HOM) effect \cite{HOM} to offer greater precision and a layer of security not afforded to classical protocols. Security is provided by employing post-selection on BB84 states for the purpose of source verification, and leveraging channel statistics to combat against asymmetric delays.

The HOM effect is a quantum interference phenomenon that occurs when two identical photons are incident upon a 50/50 beam splitter. If the two photons arrive at the beam splitter at the same time and with the same polarization, they will interfere destructively and both be routed to the same output port of the beam splitter. On the other hand, if the two photons arrive at the beam splitter with a time delay or with different polarizations, the interference will be diminished and the photons will not experience the bunching effect (see \cite{brańczyk2017hongoumandel, Moschandreou_2018, crum2025} for more details). The HOM effect can be used for clock synchronization in quantum networks by exploiting the fact that the probability of the two photons interfering depends on the time delay between them. Thus, by measuring the probability of the two photons interfering as a function of their relative arrival time at the beam splitter, it is possible to determine the relative timing of two clocks.

High‑precision quantum clock synchronization (QCS) based on HOM interference has been demonstrated with frequency‑correlated spontaneous parametric down-conversion (SPDC) photon pairs \cite{Quan_22,Liu_qcs_demonstration, Lee_19, lafler_23, Spiess_23} with security further discussed in \cite{Quan_2024}. These proof‑of‑principle experiments achieved sub‑picosecond clock difference uncertainties over tens of kilometers of fiber but can be limited by the probabilistic nature of SPDC. For example, a recent multi‑user QCS demonstration over a 75~km entanglement distribution network reported a clock difference uncertainty of about 4.5~ps and a time deviation of 426~fs after averaging for 4000~s \cite{Tang2023}. A bottleneck in such schemes is the low pair production rate of heralded single‑photon sources: commercially available SPDC modules produce only $\sim 100$–450~kHz photon pairs \cite{ThorlabsSPDC}, several orders of magnitude lower than the tens of MHz repetition rates available from pulsed laser diodes. Additionally, single-photon sources are heavily limited by transmission distance due to fiber attenuation. Consequently, high‑precision experiments based on SPDC photons are fundamentally limited in the number of useful trials.

In contrast, attenuated WCPs can provide pulse trains with repetition rates up to tens of MHz, enabling a much larger number of synchronization trials in a fixed measurement time. Coherent states contain a non‑zero probability of multi‑photon emission, which allows for greater transmission distance, but such behavior reduces the visibility of the HOM dip compared to ideal single‑photon sources. Nevertheless, the vastly increased repetition rate and larger resilience to photon loss motivate a WCP‑based protocol, which is the focus of this work.

The problem of clock synchronization involves two spatially separated parties, Alice and Bob, each possessing local clocks, Clock A and Clock B, respectively. It is assumed that the clocks are stable and have matched rates over the synchronization interval. This is justified because modern atomic clocks, such as cesium fountain standards, exhibit frequency offsets below $10^{-14}$ over day-long intervals (equivalent to $<100$ ps/day), and state-of-the-art optical lattice clocks show instabilities as low as $5\times10^{-17}/\sqrt{\tau}$ over averaging times $\tau$ (in seconds) \cite{oelker2019stability}. Thus, clock B has an unknown offset $\delta$ relative to clock A. It is also assumed that the clocks are located in stationary reference frames and the analysis applies in the non‑relativistic limit. The goal of a clock synchronization protocol is to determine $\delta$ and apply the necessary corrections to Clock B to achieve synchronization with Clock A. 

The paper proceeds in the following fashion. Section \ref{sec:2} introduces the theoretical model for clock synchronization via HOM interference, provides the coincidence probability for the WCP configuration and demonstrates how bidirectional timing allows one to eliminate unknown channel delays. Section \ref{sec:3} describes the simulation methodology and presents numerical results for path balancing, correlation analysis and the estimated clock offset under realistic detector parameters. Section \ref{sec:4} analyzes security against intercept‑resend and photon‑number splitting attacks by examining the polarization‑averaged coincidence statistics and highlighting how deviations from the expected minima reveal an eavesdropper. Finally, Section \ref{sec:5} summarizes the conclusions and suggests directions for future work. Appendix \ref{sec:A1} presents derivations of the coincidence probability for HOM interference for the weak coherent states, and Appendix \ref{sec:A2} provides the derivations for the expected coincidence probability in the free-run and post-selected instances.

\section{Theory and Methods}\label{sec:2}
Below, we outline a synchronization protocol leveraging HOM interference using WCP sources, discuss the advantages and limitations of WCPs, and emphasize the motivation for this approach. In the proposed protocol, users Alice and Bob encode BB84 polarization states \cite{BENNETT20147} onto photons using local sources and transmit them through an optical fiber. Each employs  HOM interference, where photons interfere at a 50/50 beamsplitter. Successful synchronization hinges on minimizing coincidence detection events---i.e., when photons exit separate beamsplitter ports---by tuning local path lengths. The coincidence probability depends on photon indistinguishability, which is governed by spectral overlap, polarization alignment, and relative arrival time. A minimized coincidence rate at zero relative delay ($\tau=0$) enables precise determination of the clock offset $\delta$. Note, any distinguishability in the photon inputs alters photon statistics and degrades synchronization performance.

\begin{figure}[h!]
\centering
\includegraphics[width=\columnwidth]{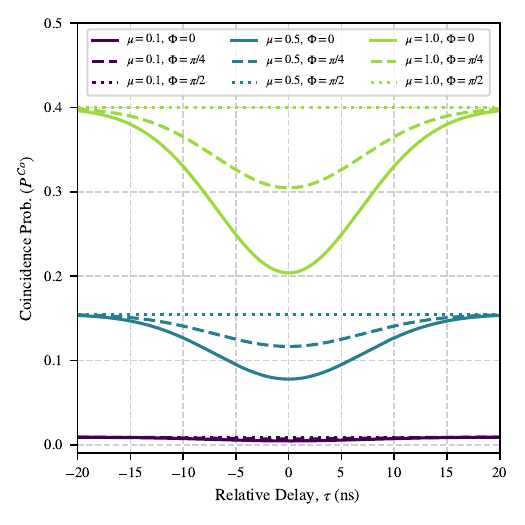}
\caption{Simulated coincidence probability (\eqref{eq:coh_coin}) as a function of relative time delay, $\tau$, for interfering WCPs. The plots show the effect of varying the mean photon number per pulse ($\mu$) and the polarization mismatch angle ($\Phi$) while holding the temporal width constant at $\sigma_{t}=10.0$ ns. As $\mu$ increases, the dip is more apparent, though relative depth is reduced, and as $\Phi$ increases from $0$ to $\pi$/2, the interference visibility is reduced, causing the dip to become shallower.}
\label{fig:mu_phi_variation}
\end{figure}

\begin{figure}[htbp]
\centering
\includegraphics[width=\columnwidth]{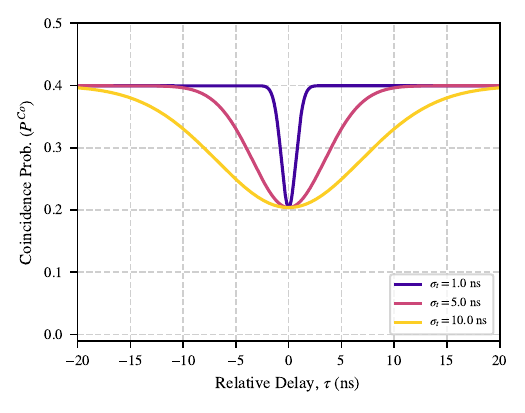}
\caption{Simulated coincidence probability versus relative time delay, $\tau$, for different spectral bandwidths, represented by the temporal width, $\sigma_t$. The mean photon number is fixed at $\mu=1.0$ and the polarization is perfectly matched. A smaller temporal width results in a narrower interference dip, demonstrating the inverse relationship between the temporal coherence of the photons and the width of the Hong-Ou-Mandel dip.}
\label{fig:sigma_variation}
\end{figure}

For WCP inputs, the coincidence probability at the beamsplitter depends on the overlap of the two coherent wave packets. When both parties send phase-randomized coherent pulses with the same mean photon number $\mu$ at the beamsplitter, the coincidence probability is given by
\begin{equation}\label{eq:coh_coin}
 P^{\mathrm{Co}} = 1 + e^{-2\mu} - 2 e^{-\mu}
 I_0\!\Big(\mu \, \cos\Phi \, e^{-\tfrac12 \sigma^2 \tau^2}\Big),
\end{equation}
where $I_0$ is the modified Bessel function of the first kind of order zero, $\tau$ is the relative delay between signal arrival times at the beamsplitter, $\sigma$ is the spectral envelope standard deviation, and $\cos\Phi$ quantifies the polarization mismatch between the pulses. This expression is derived in Appendix~\ref{sec:A1}. It shows that multi-photon contributions from the coherent state lead to a non-zero coincidence floor even at zero delay. Figure \ref{fig:mu_phi_variation} shows the coincidence probability for various average photon numbers and polarization mismatches. Figure \ref{fig:sigma_variation} shows the HOM dip for various temporal widths of the input states.

Generating photons with indistinguishable spectra at two remote sites is an experimental challenge.  In practice the central frequencies of Alice’s and Bob’s lasers must be matched to within the spectral bandwidth of the pulses (tens of MHz for 10~ns pulses) and their linewidths kept narrow relative to this bandwidth.
Achieving stable HOM interference with two independent lasers is achievable as experiments using frequency‑stabilized continuous‑wave lasers observed HOM fringes without a shared reference frame\,\cite{Kim2021}.
The reliability of the HOM effect requires the ability to independently produce identical spectral photons at each party's location.
Consequently, pulse shape mismatch, polarization misalignment, and source time jitter impact the coincidence probability and the precision of the clock synchronization procedure. Work has been done to extensively characterize mode mismatches across degrees of freedom for different realizations of HOM interference \cite{crum2025}.

\begin{figure}[t]
\centering
\includegraphics[width=0.45\textwidth]{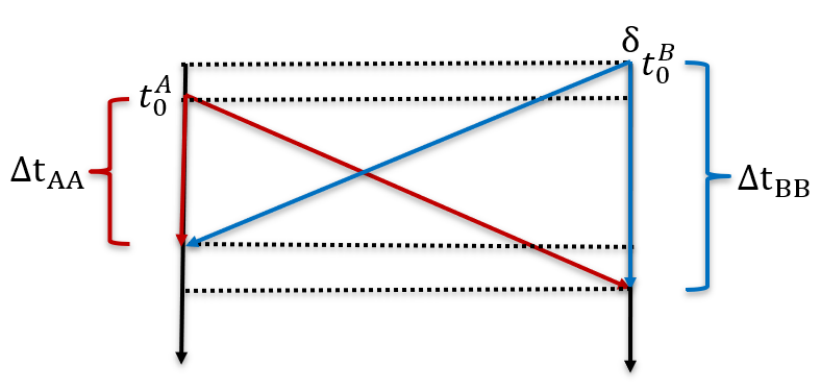}
\caption{Spacetime diagram of the two-way timing exchange between Alice (left world line) and Bob (right world line). The protocol aims to determine the unknown clock offset, $\delta$, between their local clocks. Alice sends a signal at her local time $t_0^A$ (red line), and Bob sends a signal at his local time $t_0^B$ (blue line). By using the locally measured time intervals $\Delta t_{AA}$ and $\Delta t_{BB}$, the offset can be determined while canceling the unknown propagation delay of the channel.}
\label{fig:worldline}
\end{figure}

To describe the protocol, we define the local clock times for Alice as $t^A$ and for Bob as $t^B$. We set Alice's clock as the reference. Bob's clock is assumed to run at the same rate but has an unknown, constant offset, $\delta$, relative to Alice's. Their local times are therefore related by:
\begin{equation}
 t^B = t^A + \delta
\end{equation}
The goal of the protocol is to determine this offset, $\delta$. The protocol begins when each party starts their transmission at their respective local time zero ($t_0^A = 0$ and $t_0^B = 0$). From the perspective of Alice's reference frame, she starts at $t^A_0 = 0$. Bob starts when his clock reads $t^B_0 = 0$, which corresponds to Alice's time $t^A = t^B - \delta = 0 - \delta = -\delta$.

\begin{figure*}[t]
    \centering
    \includegraphics[width=\textwidth]{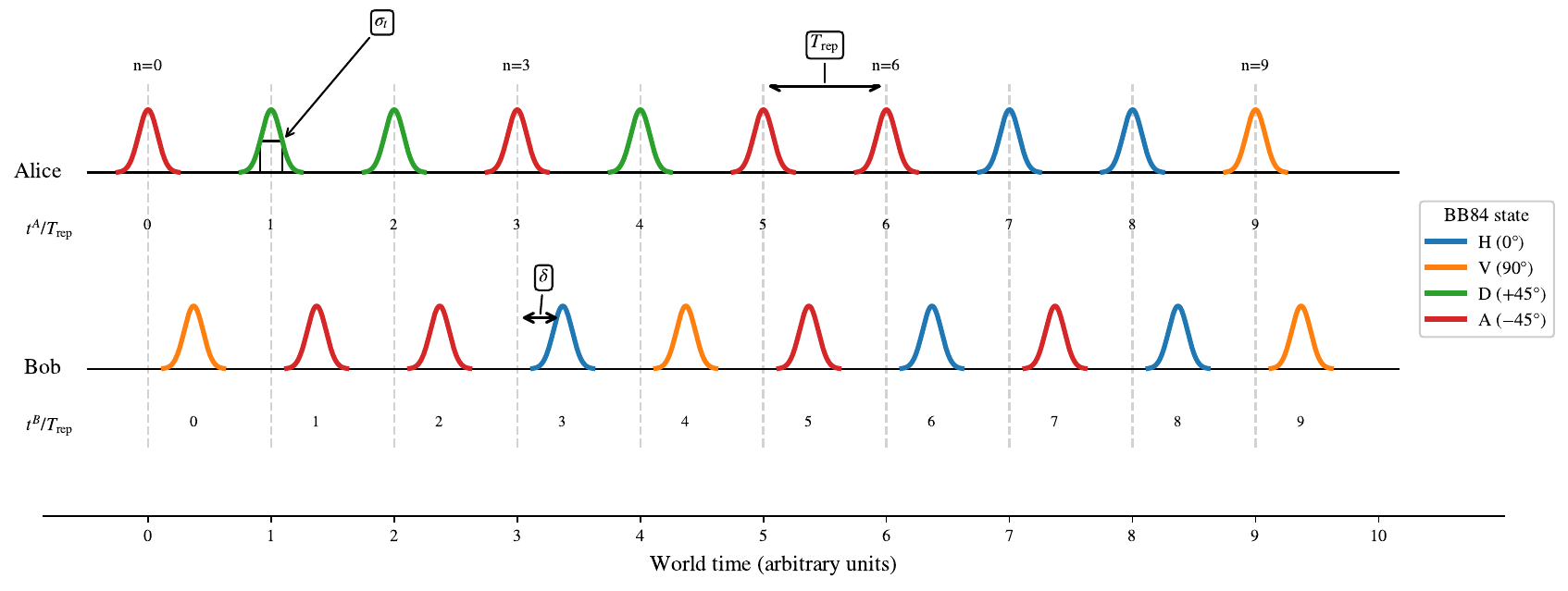}
    \caption{Schematic for Alice and Bob's synchronization signals. Both prepare pulses with randomly selected BB84 polarization state separated in time by a period $T_{\mathrm{rep}}$. Each pulse has a temporal width $\sigma_t$. The disparity in the start time is the offset $\delta$. The color of each pulse indicates the prepared BB84 polarization state.}
    \label{fig:pulse_trains}
\end{figure*}

Alice and Bob each send a train of $N$ pulses with a period $T_{\mathrm{rep}}$. Expressed in Alice's time frame, the set of emission events for each party is given by:
\begin{equation}
 \label{schedule}
 \begin{split}
     &\mathcal{T}^A=\bigl\{0,\;T_{\mathrm{rep}},\;2T_{\mathrm{rep}},\;\dotsc,\;(N-1)T_{\mathrm{rep}}\bigr\},\\
     &\mathcal{T}^B=\bigl\{-\delta,\;T_{\mathrm{rep}}-\delta,\;\dotsc,\;(N-1)T_{\mathrm{rep}}-\delta\bigr\}.
\end{split}
\end{equation}
Here $N$ denotes the total number of pulses emitted by each party during a synchronization frame and $T_{\mathrm{rep}}$ is the period of repetition between consecutive pulses. We assume that $N$ and $T_{\mathrm{rep}}$ are known and set locally by each party. See Figure \ref{fig:pulse_trains} for a depiction of the emission schedules.

First, Alice sends her pulses to Bob's lab. The signals travel a path with a propagation time of $\Delta t_{AB}$. Bob's signals travel locally to his interferometer over a path with an adjustable delay time of $\Delta t_{BB}$. Coincidence at Bob's beamsplitter for Alice's $r$-th pulse and Bob's $l$-th pulse occurs when their arrival times are identical in a common reference frame (e.g., Alice's):
\begin{equation}
 \underbrace{r \cdot T_{\mathrm{rep}} + \Delta t_{AB}}_{\text{Alice's pulse arrival}} = \underbrace{(l \cdot T_{\mathrm{rep}} - \delta) + \Delta t_{BB}}_{\text{Bob's pulse arrival}}
\end{equation}
Defining the integer index difference as $k=l-r$ and rearranging gives an expression for the clock offset:
\begin{equation}
 \delta = k\,T_{\mathrm{rep}} - \Delta t_{AB} + \Delta t_{BB}
 \label{eq:delta_T1}
\end{equation}
Experimentally, Bob identifies the correct index difference $k$ by finding the pair of pulses $(r, l)$ that produces the maximal HOM suppression. Specifically, Bob records timestamps for all detection events and constructs a two-dimensional histogram of the relative arrival times of his local and remote pulses. The correct pair corresponds to a clear dip in coincidence counts. Note, this equation still contains the unknown propagation time $\Delta t_{AB}$.

To eliminate this unknown, the roles are reversed. Bob sends his signals to Alice's lab over a path with propagation time $\Delta t_{BA}$, and Alice's signals travel locally through a delay $\Delta t_{AA}$. Coincidence for Bob's $r'$-th pulse and Alice's $l'$-th pulse occurs when:
\begin{equation}
 \underbrace{(r' \cdot T_{\mathrm{rep}} - \delta) + \Delta t_{BA}}_{\text{Bob's pulse arrival}} = \underbrace{l' \cdot T_{\mathrm{rep}} + \Delta t_{AA}}_{\text{Alice's pulse arrival}}
\end{equation}
Defining $k'=l'-r'$ gives a second expression for the clock offset:
\begin{equation}
 \delta = -k'\,T_{\mathrm{rep}} + \Delta t_{BA} - \Delta t_{AA}
 \label{eq:delta_T2}
\end{equation}
Similar to before, Alice determines the integer difference $k'$ by identifying the pair of indices $(l',r')$ that yields the maximal HOM suppression, while $\Delta t_{AA}$ is a variable local delay controlled by her.

To find the offset, we assume the channel is reciprocal, such that the propagation time is the same in both directions ($\Delta t_{AB} = \Delta t_{BA}$). By adding Eqs.\ \eqref{eq:delta_T1} and \eqref{eq:delta_T2}, the propagation delay terms cancel, and we can solve for $\delta$:
\begin{equation}
\label{eq:delta_final_rearranged}
2\delta = (k-k')T_{\mathrm{rep}} + \Delta t_{BB} - \Delta t_{AA}
\end{equation}
This gives the final expression for the clock offset in terms of locally known (measurable) quantities:
\begin{equation}
 \delta = \frac{1}{2} \left[ (k-k')T_{\mathrm{rep}} + \Delta t_{BB} - \Delta t_{AA} \right]
 \label{eq:delta_final}
\end{equation}
Again, this result relies on channel reciprocity, which holds when the channel is linear and time-invariant. For a typical 10 km link of commercial fiber (e.g., ITU-T G.655), the intrinsic propagation time asymmetry from effects like polarization-mode dispersion is less than a picosecond \cite{FSG655}. However, environmental factors like dynamic temperature gradients or adversarial tampering could introduce asymmetric delays that violate this assumption, biasing the result by a term proportional to $\Delta t_{BA} - \Delta t_{AB}$. Such non-reciprocity could be compensated for by pre-calibrating the channel or by using additional bidirectional reference pulses.

\begin{figure}[htb!]
    \centering
    \includegraphics[width=1.0\linewidth]{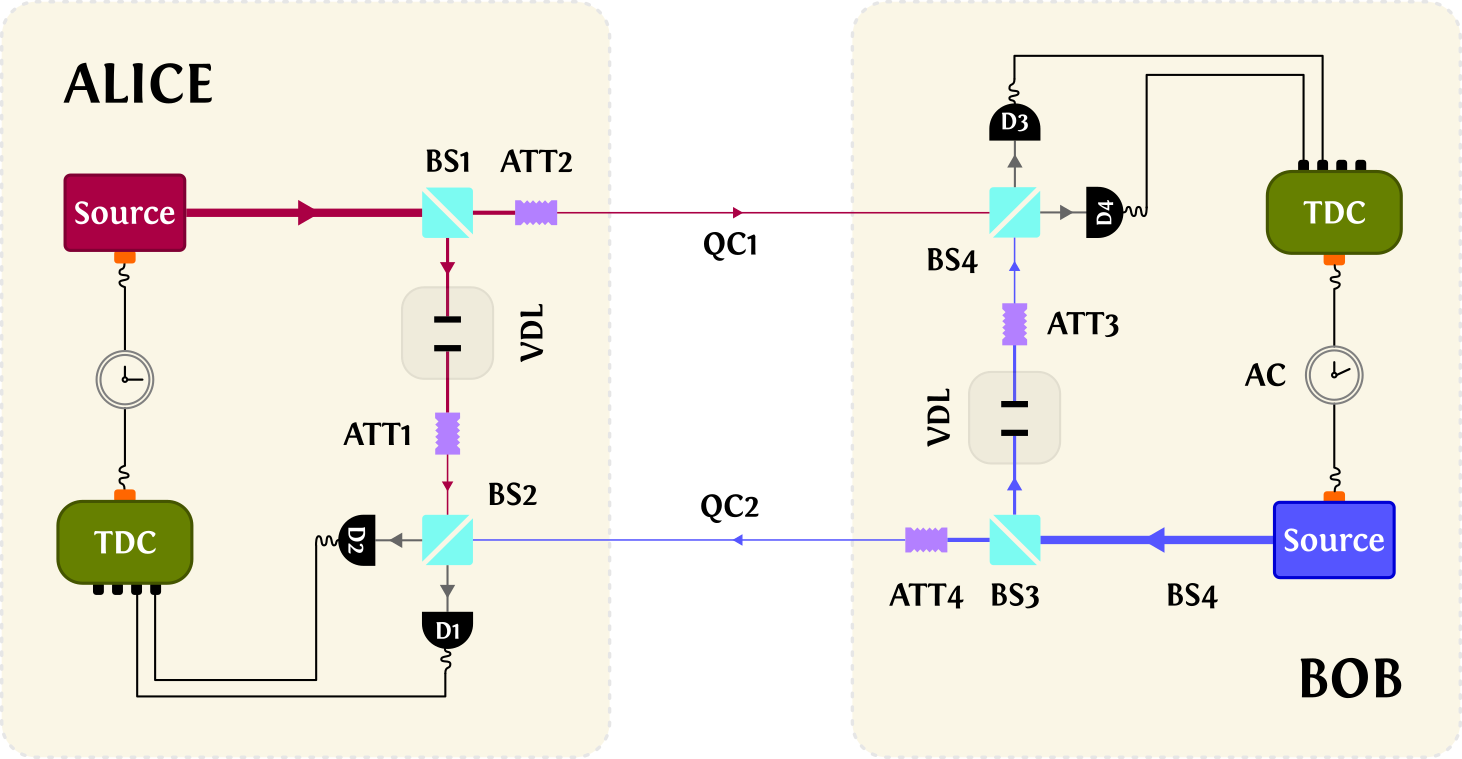}
    \caption{Schematic of the experimental protocol. Alice and Bob employ symmetric setups. The \textbf{Source} module consists of a classical communication-band laser equipped with intensity and polarization modulators to generate bright pulses with controllable polarization states. At Alice’s side, beam splitter 1, \textbf{BS1}, divides the pulses: one copy remains local, while the other is transmitted to the network. Variable optical attenuators, \textbf{ATT1} and \textbf{ATT2}, reduce the pulse intensity to the desired photon level ($\mu$). The pulse attenuated by ATT2 is sent through the quantum channel, \textbf{QC1}, to Bob. At \textbf{BS2}, Alice’s locally attenuated pulses interfere with those arriving from Bob’s side via the network. Single-photon detectors, \textbf{D1} and \textbf{D2}, register detection events following the interference. A variable delay line, \textbf{VDL}, compensates for the relative phase difference between the paths. Detection events are time-tagged and correlated using a time-to-digital converter, \textbf{TDC}. A precision atomic clock, \textbf{AC}, provides timing discipline for the modulators and the TDC, minimizing drift among the high-speed electronic components.}
    \label{fig:protocol_setup}
\end{figure}

Altogether, the pulse index differences ($k, k'$) are found through post-selection by locating the HOM dip, and the local delays $\Delta t_{AA}$ and $\Delta t_{BB}$ are known from the settings of the local delay lines. Equation \eqref{eq:delta_final} thus allows for a direct calculation of the clock offset. See Figure \ref{fig:protocol_setup} for a depiction of the physical realization of this setup.

\section{Results}\label{sec:3}
For a simulated true clock offset of  $\delta=230.456 \,\mathrm{ns}$, our protocol determines an estimate of the clock offset $\hat{\delta}=230.462 \pm 0.027\, \mathrm{ns}$ giving an accuracy of $6.205$ $\mathrm{ps}$ and a standard error of $26.71$ $\mathrm{ps}$ under the conditions summarized in Table~\ref{tab:sim_params}. The simulation assumes telecom-band ($1550\, \mathrm{nm}$) Gaussian pulses with a temporal width of $10.0\, \mathrm{ns}$, a $10 \, \mathrm{MHz}$ repetition rate, $10\, \mathrm{km}$ fiber link ($2\, \mathrm{dB}$ attenuation), $85\%$ detector efficiency, and $150\, \mathrm{ps}$ detector timing jitter. To achieve an effective mean photon number of $\mu=1.0$ at the beamsplitter, an initial mean photon number of $\mu_{\text{input}} \approx 1.86$ was used at the source to compensate for these channel and local losses. We obtained this value by simulating the bidirectional protocol using a series of frames, where each run consisted of $N=10^5$ pulses per variable delay line (VDL) setting.

\begin{table}[ht!]
\centering
\caption{Simulation parameters used for the offset precision histogram at temporal width $10\,$ns and effective mean photon number $\mu=1.0$.}
\label{tab:sim_params}
\begin{tabular}{l c c}
\hline
\textbf{Parameter} & \textbf{Symbol} & \textbf{Value} \\
\hline
Number of pulses (per trial) & $N$ & $100{,}000$ \\
Repetition rate & $f_{\mathrm{rep}}$ & $10\,\mathrm{MHz}$ \\
Pulse period & $T_{\mathrm{rep}}$& $100\,\mathrm{ns}$ \\
Mean photon number & $\mu$ & $1.0$ \\
Temporal width (FWHM) & $\sigma_{t}$& $10\,\mathrm{ns}$ \\
Detector jitter (FWHM) &  -& $150\,\mathrm{ps}$ \\
Propagation delay&  $\Delta t_{AB},\, \Delta t_{BA}$& $5.0\times 10^{-5}\,\mathrm{s}$ \\
 VDL Resolution& -&$180.0\, \mathrm{ps}$\\
True offset & $\delta_{\mathrm{true}}$ & $230.456\,\mathrm{ns}$ \\
\end{tabular}
\end{table}
The simulation methodology for each direction consists of a two-stage process to determine the optimal local delay and the correct pulse index offset. 

\begin{figure}[htbp]
    \centering
    \includegraphics[width=1.0\linewidth]{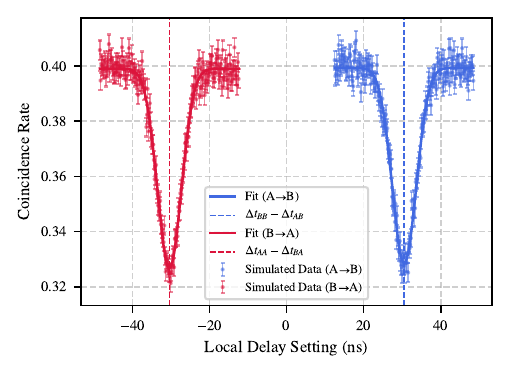}
    \caption{Results of the simulated path-balancing measurement for bidirectional clock synchronization via WCP. The plot displays the post-selected coincidence rate as a function of the local electronic delay setting applied at the receiving station, with the channel delay removed via the reciprocity. Data points with binomial error bars and their corresponding weighted inverted Gaussian fits are shown for both the Alice to Bob (blue) and Bob to Alice (red) channels. The dashed vertical lines mark the minima of the fitted HOM dips representing the optimal local delays.}
    \label{fig:hom_sim}
\end{figure}

\begin{figure}[htbp]
    \centering
    \includegraphics[width=1.0\linewidth]{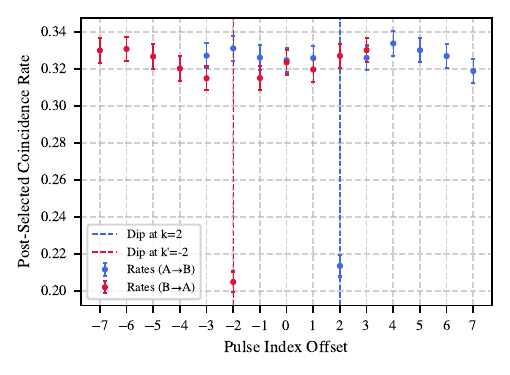}
    \caption{Results of the simulated correlation analysis used to determine the integer pulse index offset. The plot shows the coincidence rate as a function of the integer offset applied during post-processing of the arrival time data. Results are shown for both the Alice to Bob (blue) and Bob to Alice (red)  directions after applying the optimal local delays. A distinct dip in coincidences occurs only at the correct offset that aligns the corresponding pulses. }
    \label{fig:path_balance}
\end{figure}

In the first stage, a path-balancing scan is performed, see Figure \ref{fig:hom_sim}. The local delay line at the receiving station is swept across a range centered on an expected value determined from the coarse timing difference between local and remote emission schedules. For each delay setting, coincidence events are generated probabilistically according to the theoretical coincidence probability given in Eq.~\eqref{eq:coh_coin}, which accounts for both the relative arrival time and the polarization mismatch of each pair. The outcome of this Monte Carlo procedure is a coincidence rate, and the coincidence rate as a function of delay forms the HOM dip, which is fit to an inverted Gaussian model using a weighted least-squares algorithm. The statistical uncertainties provide the weights for the fit, and the covariance matrix of the fit yields the standard error of the fitted parameters. The fitted center of the Gaussian specifies the optimal local delay, while the standard error of this parameter quantifies the uncertainty in the delay estimate.

In the second stage, a correlation analysis is applied as a post-processing step on the event record for the point nearest the optimal delay obtained in the previous step. The coincidence rate is computed as a function of the integer pulse index offset ($k$ or $k'$), restricted to the subset of polarization-matched BB84 states. Each coincidence rate is assigned an error bar assuming binomial statistics. The correct pulse index offset is identified as the integer value that minimizes the post-selected coincidence rate, see Figure \ref{fig:path_balance}. 

Finally, the four measured quantities $\Delta t_{AA}$, $\Delta t_{BB}$, $k$, and $k'$ are substituted into Eq.~\eqref{eq:delta_final} to calculate the clock offset. The total uncertainty is obtained by propagating the fit errors on $\Delta t_{AA}$ and $\Delta t_{BB}$. Table~\ref{tab:sim_results} summarizes the numerical values obtained from the simulation. The provided uncertainties for the fitted local delays correspond to one standard deviation confidence intervals from the covariance of the inverted Gaussian fits. The final clock offset precision is reported as the propagated 1$\sigma$ uncertainty.

\begin{table}[ht]
\centering
\caption{Simulation results for the bidirectional protocol. The local delays are determined from inverted Gaussian fits to the HOM dips, while the pulse index offsets are identified from the correlation analysis. The final clock offset $\hat{\delta}$ is computed from Eq.~\eqref{eq:delta_final}.}
\label{tab:sim_results}
\begin{tabular}{l c}
\hline
\textbf{Quantity} & \textbf{Value} \\
\hline
Optimal delay $\Delta t_{BB}$ (A$\to$B) &  $50030.513 \pm 0.038\, \mathrm{ns}$\\
Optimal delay $\Delta t_{AA}$ (B$\to$A) &   $49969.588 \pm 0.037\, \mathrm{ns}$\\
Pulse index offset $k$ (A$\to$B) & $2$ \\
Pulse index offset $k'$ (B$\to$A) & $-2$ \\
Final clock offset $\hat{\delta}$ & $230.462 \pm 0.027$ ns\\
True clock offset $\delta_{\mathrm{true}}$ & $230.456$ ns \\
\hline
 Accuracy $|
\delta_{\mathrm{true}} - \hat{\delta}|$&$6.205\, \mathrm{ps}$\\
\end{tabular}
\end{table}

\section{Protocol Security}\label{sec:4}

In our protocol each party independently prepares weak coherent pulses with polarizations chosen uniformly at random from the four BB84 states. Because the polarization choice is revealed only after the quantum exchange, any malicious modification of the pulse stream will necessarily disturb the interference statistics.  Classical post‑selection, retaining only those events in which Alice and Bob used the same basis, therefore provides a mechanism to verify that the interfering pulses originated from legitimate transmitters.  The randomness of the basis choices means that the relative polarization mismatch $\Phi$ between Alice and Bob is a random variable taking values $\Phi=0,\, \pi/4$ and $\pi/2$ with probabilities 0.25, 0.5 and 0.25, respectively.  Averaging the coincidence probability $P^{\mathrm{Co}}(\tau=0)$ over these mismatches yields Eq.~(\ref{expect_coin}), which establishes the expected coincidence floor for uncorrelated polarizations.  
\begin{equation}\label{expect_coin}
    \langle P^{\mathrm{Co}}(\tau=0) \rangle_\Phi = 1 + e^{-2\mu} - \frac{e^{-\mu}}{2} \Bigl[1 + I_0(\mu) + 2 I_0 (\tfrac{\mu}{\sqrt{2}} )\Bigr]
\end{equation}
Conversely, when Alice and Bob post‑select on matched polarizations ($\Phi=0$), the HOM interference is maximal and the coincidence probability at zero delay reduces to Eq.~(\ref{ps_coin}), 
\begin{equation}\label{ps_coin}
    P^{\mathrm{Co}}(\Phi=0, \tau=0) = 1 + e^{-2\mu} - 2 e^{-\mu} I_0(\mu)
\end{equation}
which is the minimum achievable with weak coherent pulses.  Monitoring the post‑selected coincidence rate relative to these theoretical benchmarks allows the parties to authenticate the source and to detect adversarial disturbances.

The primary active threat considered here is the intercept–resend (IR) attack, in which an eavesdropper Eve attempts to measure the polarization of each incoming pulse and to retransmit a new pulse toward the receiver.  Because the interferometer is sensitive to both the arrival time and the polarization, measuring and re‑preparing the state necessarily introduces bit‑flip ($X$) and phase‑flip ($Z$) errors.  In the rectilinear basis HOM visibility is robust against $Z$ errors but degraded by $X$ errors, while in the diagonal basis the situation is reversed.  An intercept–resend strategy therefore produces a mixture of basis errors that elevates the coincidence rate in both bases simultaneously.  

\begin{figure}
    \centering
    \includegraphics[width=1\linewidth]{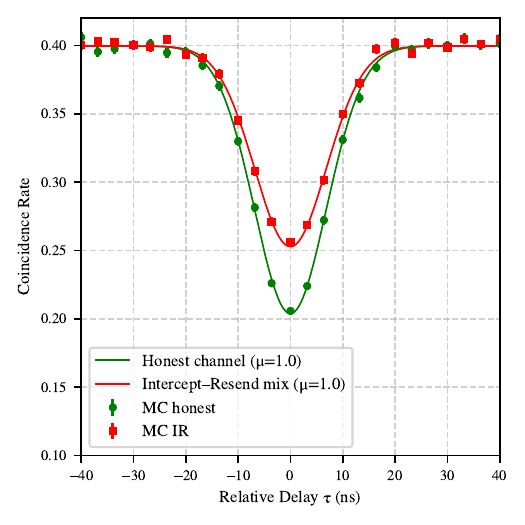}
    \caption{HOM dip for $\mu=1.0$ comparing the honest channel (solid green) to an intercept--resend (IR) mixture with a $25\%$ matched-basis error (solid red). The IR attack elevates the coincidence floor above the theoretical minimum. Symbols show Monte Carlo simulation results with binomial error bars ($20{,}000$ trials per point).}
    \label{fig:honest_vs_IR}
\end{figure}

\begin{figure}
    \centering
    \includegraphics[width=1.0\linewidth]{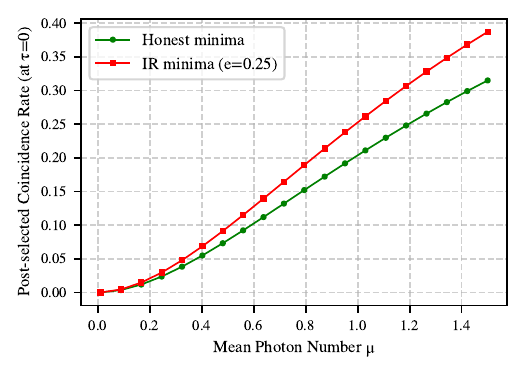}
    \caption{Post-selected coincidence probability at zero delay ($\tau=0$) as a function of mean photon number $\mu$. Green circles correspond to the honest channel, while red squares show the elevated minima induced by an intercept--resend (IR) attack with 25\% error rate. The systematic separation between the two curves provides a quantitative signature of eavesdropping.}
    \label{fig:attack_vs_mu}
\end{figure}

Quantitatively, when Eve measures in a random basis and resends her guess, the post‑selected coincidence probability no longer approaches the minimum value of Eq.~(\ref{ps_coin}) but instead approaches the higher averaged value of Eq.~(\ref{expect_coin}).  Figures\,\ref{fig:honest_vs_IR} and\,\ref{fig:attack_vs_mu} illustrate this effect.  In Fig.~\ref{fig:honest_vs_IR} the HOM dip for $\mu=1$ is plotted for both an honest channel (green) and an IR mixture with a 25\,\% matched‑basis error rate (red); the latter exhibits a noticeably higher coincidence floor than the honest channel.  Figure\,\ref{fig:attack_vs_mu} shows the post‑selected coincidence probability at zero delay as a function of mean photon number. Again, the IR attack raises the minima (red squares) above the honest threshold (green circles).  Because random channel imperfections tend to affect one basis more than the other, simultaneously elevated coincidence rates in both bases provide a distinctive signature of an intercept–resend attack.  Alice and Bob can therefore set a statistical threshold based on Eqs.~(\ref{expect_coin}) and~(\ref{ps_coin}); if the measured post‑selected coincidence rate exceeds this threshold they abort the synchronization and infer the presence of an eavesdropper.

Weak coherent sources occasionally emit multi‑photon pulses, opening the door to photon–number splitting (PNS) attacks in which Eve nondestructively measures the photon number, diverts one photon to learn the polarization and forwards the remainder.  In quantum key distribution such an attack can reveal key bits undetected, but in our synchronization protocol the encoded polarization is used solely for source authentication; the arrival times of the pulses carry the timing information.  Consequently, a PNS attack could assist Eve in guessing the basis choices and circumventing the post‑selection, but it cannot directly extract the clock offset.  Nonetheless, to preserve source authentication and to detect PNS attempts it is prudent to verify the photon‑number statistics of the channel.  A standard technique from QKD is the decoy–state method, first proposed by Hwang and rigorously analyzed by Lo, Ma and Chen \cite{LoMaChen2005}. The key idea is for Alice to intersperse the signal pulses with additional ``decoy'' pulses of different intensities.  Because Eve cannot distinguish a decoy pulse from a signal pulse except through its photon number, the yields and error rates of decoy and signal states must depend only on the photon number and not on the intensity distribution.  By measuring the detection probabilities (yields) for several intensities, Alice and Bob can bound the fraction of single‑photon events and detect discrepancies indicative of a PNS attack.  

In the context of clock synchronization, incorporating a small set of vacuum and weak‑intensity decoy pulses would allow the legitimate parties to estimate the photon‑number distribution of the channel and confirm that the observed coincidence statistics are consistent with the expected Poissonian distribution.  Such a decoy‑state check complements the post‑selection authentication and renders PNS attacks ineffective.  A detailed implementation of decoy states for synchronization is left for future work, but the underlying technique is well established in QKD.  In their seminal paper Lo et al. emphasize that decoy states are additional intensity levels used solely to detect eavesdropping, while the standard signal states carry the key information \cite{LoMaChen2005}; this distinction naturally aligns with our separation between timing information and source authentication. Other examples of QKD attacks include Trojan-horse attacks on Alice’s or Bob’s preparation modules, wavelength-dependent manipulations of dispersion or beam-splitter ratios, and detector side-channel vulnerabilities. These techniques are well documented in the QKD literature \cite{sajeed2017trojan,fei2018wavelength,sun2022review,pantoja2024sidechannel}, though here their effect would be to shift the location of the HOM minimum or alter coincidence statistics rather than to leak classical information.

While a complete adversarial analysis lies beyond the scope of this work, standard countermeasures from QKD remain applicable. Optical isolators limit Trojan-horse probes, spectral filtering suppresses wavelength-based channel manipulations, detector monitoring helps identify anomalous behavior, and decoy-states circumvent PNS attacks. For field deployment, these defenses are necessary to ensure that the synchronization offset cannot be systematically manipulated. Importantly, the theoretical framework developed here remains valid under such protections; future work should extend this analysis to quantify robustness against explicit device-level attacks.

\section{Conclusion}\label{sec:5}
Through the use of WCPs, our simulated protocol for clock synchronization allows for determination of the offset between two spatially separated clocks to a precision limited by the temporal properties of the photon source and the specified tolerance for HOM path balancing. The protocol relies on the strong assumption of a symmetric communication channel with a constant propagation delay. This corresponds to the reciprocity assumption used in the theoretical derivation, namely that the forward and backward propagation times through the fiber are identical. In real networks this equality can be violated by polarization‑mode dispersion, chromatic dispersion or temperature‑dependent refractive index changes. If the channel is subject to environmental variations on the timescale of the sync or a malicious adversary inserts an extra delay in one direction, the reciprocity assumption fails and the offset extracted from Eq.\ \eqref{eq:delta_final} will be biased. To mitigate this, Alice and Bob can perform regular calibrations using probe pulses sent in both directions at a different wavelength, estimate the difference $\Delta t_{AB}-\Delta t_{BA}$ and subtract it from Eq.\ \eqref{eq:delta_final}.

Beyond periodic calibration, Alice and Bob can continuously monitor and compensate dynamic asymmetries by interleaving bright classical calibration pulses at a wavelength outside the quantum channel (e.g., 1310~nm).  By measuring the round‑trip delays of these pulses and comparing the forward and backward travel times, the parties can estimate $\Delta t_{AB}-\Delta t_{BA}$ in real time and update their local delays $\Delta t_{AA}$ and $\Delta t_{BB}$ accordingly.  Alternatively, a classical two‑way time transfer protocol can be run in parallel with the quantum exchange to actively track the propagation asymmetry and feed forward corrections into the HOM timing analysis.  These strategies enable deployment over installed fiber links subject to temperature or mechanical fluctuations without interrupting the quantum protocol.

The protocol utilizes identical spectral photons with randomly selected polarizations corresponding to the four BB84 protocol states. Alice and Bob exchange these photons to balance separate HOM interferometers which allows them to deduce the physical time delay between. Many other QCS techniques utilize SPDC sourced photons which have limited pair production rates when compared to the photon pulse production rates of attenuated coherent sources \cite{Lee_symm_sync, Lee:22}. Utilizing WCPs allows for synchronization over larger distances by better overcoming channel loss. An important practical advantage of WCP sources is that the mean photon number $\mu$ can be tuned over orders of magnitude to compensate for fiber attenuation.  The HOM interference visibility depends on the effective photon number arriving at the interferometer, not on the absolute power launched into the channel.  As long as the pulses are attenuated such that the mean number of photons at each interferometer remains $\lesssim 1$, the characteristic HOM dip and timing relations are preserved.  This freedom means that Alice and Bob can increase $\mu$ at the transmitters to offset fiber loss and extend the transmission distance without sacrificing the ability to observe the HOM dip. By contrast, SPDC sources are constrained by fixed pair production rates and saturate at long distances.  Thus, WCP‑based synchronization inherently affords a transmission‑distance advantage over single‑photon schemes.

Our protocol offers source verification to clock synchronization through the process of post-selection. If Alice and Bob are unable to observe strong HOM visibility in the post-selected basis, they can infer the presence of an eavesdropper performing an intercept and resend attack. Practically, post‑selection is implemented by comparing the polarization settings chosen by Alice and Bob for each pulse pair and retaining only those events where the bases match. This requires classical communication of the basis choices after the quantum exchange, analogous to sifting in BB84. The trade‑off is a large portion of the raw detection events are discarded, reducing the effective synchronization rate by the same factor. However, retaining only the matched bases improves interference visibility and thus maximizes the precision of the offset estimate.

The immediate next step is an experimental demonstration of WCP-based HOM synchronization over 10–50 km of deployed fiber. With realistic detectors (timing jitter $\approx$ 150 ps, efficiency $\approx$ 85\%), our protocol targets sub-100 ps accuracy in metro-scale links and sub-10 ps accuracy in laboratory-scale testbeds after averaging. We envision this method serving as the authenticated timing backbone of quantum repeater networks, where secure and precise clock synchronization will be essential for entanglement swapping, distributed quantum computing, and networked sensing.

\acknowledgments

This material is based upon work supported by the  Department of Energy, Office of Science, Office of Advanced Scientific Computing Research, through the Quantum Internet to Accelerate Scientific Discovery Program under Field Work Proposal 3ERKJ381.
We acknowledge support by the National Science Foundation under grant DGE-2152168.

\bibliography{qcs_sources}

\appendix
\section{Coincidence Probabilities for Matched Gaussian Spectra with Polarization Mismatch}
\label{sec:A1}
We adopt the multi-mode HOM framework with polarization and spectro-temporal overlaps
\begin{align}
\cos\Phi &= \big|\hat\epsilon_A^\ast\!\cdot\!\hat\epsilon_B\big|,\nonumber\\
\cos\Theta &= \Big|\int_{-\infty}^{\infty} d\omega\, \varphi_A^\ast(\omega)\,\varphi_B(\omega)\Big|.
\end{align}
For normalized Gaussians with matched center $\omega_0$ and bandwidth $\sigma$,
\begin{align}
&\varphi_A(\omega)=\frac{1}{(2\pi\sigma^2)^{1/4}}
\exp\!\left[-\frac{(\omega-\omega_0)^2}{4\sigma^2}\right],\nonumber\\
&\varphi_B(\omega)= \varphi_A(\omega)\,e^{i\omega\tau},
\end{align}
the spectro-temporal overlap evaluates to
\begin{equation}
\cos\Theta=\exp\!\left(-\tfrac12\sigma^2\tau^2\right).
\end{equation}
Define
\begin{equation} \label{eq:overlap}
S = \cos\Phi\,\cos\Theta = \cos\Phi\,e^{-\tfrac12\sigma^2\tau^2}.
\end{equation}
For phase-randomized coherent states with means $\mu_A,\mu_B$,
\begin{align}
P^{\mathrm{co}}_{\mathrm{WCP}\times\mathrm{WCP}}
=1&+e^{-\mu_A-\mu_B} -e^{-\mu_A-\mu_B}\nonumber\\
&\times\left(e^{\mu_A R+\mu_B T}+e^{\mu_A T+\mu_B R}\right)
\nonumber\\
&\times I_0\!\big(2\sqrt{\mu_A\mu_B\,RT}\,S\big),
\label{eq:wcp_wcp_general}
\end{align}
with $I_0$ the modified Bessel function.
Insert $S$ into \eqref{eq:wcp_wcp_general} and for $T=R=\tfrac12$ with $\mu_A=\mu_B=\mu$,
\begin{equation}
P^{\mathrm{co}}_{\mathrm{WCP}\times\mathrm{WCP}}
=1+e^{-2\mu}-2e^{-\mu} I_0\!\Big(\mu\,\cos\Phi\,e^{-\tfrac12\sigma^{2}\tau^{2}}\Big).
\end{equation}

\section{Security Analysis with BB84 State Preparations}
\label{sec:A2}
In this appendix we evaluate the expected Hong--Ou--Mandel (HOM) interference minima and maxima when Alice and Bob each prepare BB84 polarization states uniformly at random from the set
\(\{|H\rangle,|V\rangle,|+\rangle,|-\rangle\}\).
The coincidence probabilities depend on the indistinguishability parameter \eqref{eq:overlap}. At the HOM minimum (\(\tau=0\)) we have \(\cos\Theta=1\) and thus \(S=\cos\Phi\).

When two BB84 states are chosen independently and uniformly at random, their polarization overlaps fall into three categories:
\[
|\langle a|b\rangle|^2 = 
\begin{cases}
1 & \text{with probability } \tfrac14 \\
0 & \text{with probability } \tfrac14 \\
\tfrac12 & \text{with probability } \tfrac12 \\
\end{cases}
\]
corresponding to identical, orthogonal, and diagonal mismatched states respectively.
Therefore
\(\mathbb{E}[\cos^2\Phi] = \tfrac12\).
These averages determine the expected HOM minima when sources are driven with BB84-prepared photons.

For two weak coherent pulses with mean photon numbers \(\mu_A=\mu_B=\mu\),
\[
P_{\min}(\tau{=}0)=1+e^{-2\mu}-2e^{-\mu} I_0\!\big(\mu\cos\Phi\big),
\]
where \(I_0\) is the modified Bessel function.
Averaging over the BB84 ensemble with weights \(\{1/4,1/4,1/2\}\) for \(\cos\Phi\in\{1,0,1/\sqrt2\}\),
\[
\mathbb{E}[P_{\min}]
=1+e^{-2\mu}
-2e^{-\mu}\!\left[
\tfrac14 I_0(\mu)+\tfrac14 I_0(0)+\tfrac12 I_0\!\Big(\tfrac{\mu}{\sqrt2}\Big)
\right]\ .
\]

When Alice and Bob post-select polarization-matched inputs (\(\cos\Phi=1\)), the visibility is
\[
V=\frac{P_{\max}-P_{\min}}{P_{\max}},
\]
with \(P_{\max}=P(\tau\!\to\!\infty,S\to 0)\) and \(P_{\min}=P(\tau=0,S=1)\).

\begin{align}
P(\tau)&=1+e^{-2\mu}-2e^{-\mu}I_0(\mu S),\nonumber\\
P_{\max}&=1+e^{-2\mu}-2e^{-\mu},\nonumber\\
P_{\min}&=1+e^{-2\mu}-2e^{-\mu}I_0(\mu),\nonumber
\end{align}
\[
 V=\frac{I_0(\mu)-1}{2 \sinh^2 \frac{\mu}{2}} \ .
\]
For WCP sources the averaged minimum quantifies the residual coincidence floor arising from random BB84 inputs.  Post‑selected matched cases give the upper bound on the achievable HOM visibility.

\end{document}